# Passive Mechanical Vibration Processor for Wireless Vibration Sensing

*Dajun Zhang[1], Akhil Polamarasetty[2], Muhammad Osama Shahid[1], Bhuvana Krishnaswamy[1], Chu Ma[1]\**

[1]*Department of Electrical and Computer Engineering, University of Wisconsin–Madison, Madison, WI 53706, USA*

[2]*Department of Computer Sciences, University of Wisconsin–Madison, Madison, WI 53706, USA*

*\*E-mail: chu.ma@wisc.edu*

## Abstract

Real-time, low-cost, and wireless mechanical vibration monitoring is necessary for industrial applications to track the operation status of equipment, environmental applications to proactively predict natural disasters, as well as day-to-day applications such as vital sign monitoring. Despite this urgent need, existing solutions, such as laser vibrometers, commercial Wi-Fi devices, and cameras, lack wide practical deployment due to their limited sensitivity and functionality. In this work, we propose and verify that a fully passive, resonance-based vibration processing device attached to the vibrating surface can improve the sensitivity of wireless vibration measurement methods by more than 10 times at designated frequencies. Additionally, the device realizes an analog real-time vibration filtering/labeling effect, and the device also provides a platform for surface editing, which adds more functionalities to the current non-contact sensing systems. Finally, the working frequency of the device is widely adjustable over orders of magnitudes, broadening its applicability to different applications.

# Introduction

Real-time mechanical vibration sensing is needed to capture the rich information in a variety of industrial applications, health care applications, natural environmental monitoring and prevention applications[1]. For example, vibration monitoring is critical for the prediction of natural disasters such as typhoons, earthquakes, and avalanche calamities, and for meteorological observation and geological survey[2]. Vibrations induced by human breathing and heartbeat are important vital signals for health monitoring[3]. Vibration monitoring also provides in-situ and non-destructive tools for diagnosing the structural health of vehicles[4], industrial equipment[5], buildings, and public infrastructures[6]. Furthermore, monitoring the resonances in ultra-high frequency mechanical vibration systems serves as an important mechanism for high-sensitivity nanoelectronics sensors[7–9]. This vast range of applications has motivated researchers to explore and develop various non-contact and contact vibration measurement techniques[10,11].

Contact-based methods typically attach sensors to the vibrating surface and hence render higher accuracy; vibration displacement or acceleration measurements using strain type[12,13], piezoelectric[14,15], or electrokinetic[16] sensors are some of the well-known approaches. However, their high accuracy comes at the cost of their need for on-board signal processing that in turn adds to the overall cost for infrastructure, including data acquisition, high quality data transmission, dedicated power supply[17]. Non-contact methods, on the other hand, are relatively less expensive as they allow for measurement and acquisition of mechanical vibrations without attaching sensors to the vibrating surfaces. There are long existing non-contact approaches based on electromagnetic eddy current[18] and capacitance measurements[19] that measure close to the vibration surfaces, optical interference based approaches that require cumbersome and expensive laser vibrometry setups among others[20]. Recent methods have proposed the use of cameras[17,21–23], ultrasound[24], frequency identification (RFID) and Wi-Fi[25–27] as more flexible alternatives to the above identified non-contact methods. They aim to capture vibration signals at a lower cost using low complexity algorithms, in turn broadening its application with the possibility of integrating vibration sensing with Internet of Things networks[28].

While the newly developed non-contact measurement technologies based on video image processing, or doppler effect of Electromagnetic (EM) Signals (Wi-Fi/RFID/Ultrasound) are promising due to their lower cost, there is still a long way to go for their practical deployments due to a key limitation: poor sensitivity, defined as the smallest vibration amplitude that can be detected. For instance, it is extremely challenging for EM-based sensing system to detect a vibration with an amplitude much smaller than one wavelength of the electromagnetic wave. In the case of RFID, one wavelength would be around 30 cm; in the case of Wi-Fi, it would be between 6 cm and 12 cm; this limitation is because of the nearly undetectable phase change generated by the vibration amplitude to be measured. Similarly, in camera-based vibration sensing, the sensitivity is decided by the number of photosensitive sensors on chip and numerical aperture. The higher the desired sensitivity, the higher is the hardware cost, contradicting the motivation of low-cost, non-contact measurement techniques.

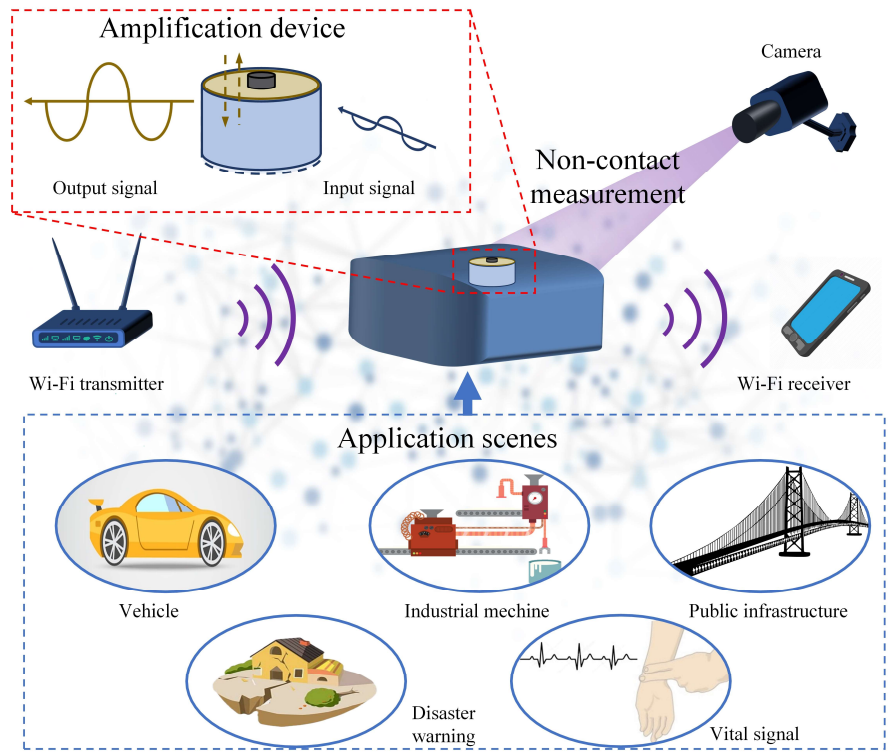

**Fig. 1. Illustration of the passive vibration amplification device and its applications.**

In this work, we propose and develop a hybrid mechanical vibration monitoring method that offers a high sensitivity to non-contact systems using our novel, passive mechanical device that amplifies the vibration of the surface of interest. The proposed passive device amplifies the vibration observed, which is then captured by a low-cost, remote data acquisition system such as camera, ultrasound, Wi-Fi, or RFID devices based on the application. The flexible design of the passive device renders it easy to tune, making it a general-purpose amplification device for a variety of applications.

Figure 1 illustrates our proposed system; the vibration processing device (Zoomed in Fig. 1 top left) is attached to the surface of interest (whose vibrations are to be monitored continuously). Its passive amplification makes it battery-free, eliminating the need for dedicated power supply that is typical of contact-based methods. The proposed device consists of a mass-loaded membrane mounted on the top of a rigid shell support[29–32]. When the device is attached or fixed on to the surface of interest, its rigid shell will vibrate along with the subject (surface), which will cause the movement of the top membrane. When the monitored subject/surface vibrates with a frequency close to the eigenfrequency of the device, the device will be excited to a mechanical resonance state. As a result, the vibration magnitude of the membrane will be much higher than that of the monitored surface for this frequency. The resonance-induced vibration magnitude amplification is exploited to enhance the vibration measurement sensitivity; it also works as a physical layer label to help source identification in non-contact vibration measurement. In our paper, we use two low-cost data acquisition systems, a Wi-Fi device and a camera, to capture the vibration amplitude that is amplified by our passive mechanical device. In the case of Wi-Fi, our proposed device's amplification improves the doppler shift of electromagnetic signals, which in turn improves the sensitivity of the system, while in case of camera video, the amplification improves the frame-to-frame variance.

The resonance frequency at which the device produces the largest amplification can be adjusted within a broad frequency range that spans several orders of magnitude (more details in design and characterization of the vibration amplification device), making our device applicable in a broad range of vibration sensing applications, such as human vital sign sensing[25,33,34] and natural disaster monitoring[35] usually below tens of hertz, structure health inspection[17,18,20,21,36–42] in the range of hertz to hundreds of hertz, and material characterization in MHz and even GHz range[43]. The mass-loaded membrane produces a resonance frequency that has a much larger wavelength compared to the membrane dimensions[44], which reduces the overall device size and hence offers flexible deployment options for a given application, especially in lower frequency range (below tens of Hz). In addition to its amplification, our device offers analog vibration signal processing functions that can be used for labelling the vibration subjects in a multi-user environment or pre-filtering the vibration signals (more details in design and characterization of the vibration amplification device). The device can also serve as a platform for surface modifications to add more functionalities in Wi-Fi- and camera- based systems, such as directional sensing capabilities for Wi-Fi sensing by attaching a Wi-Fi reflector or enabling camera-based sensing in dark environment by adding florescent dyes to the device surface.

Our work pioneers the development and application of passive vibration processing devices that improve the sensitivity and accuracy of wireless vibration measurements such as Wi-Fi and camera-based vibration sensing. Our device is purely passive[45] and does not require an additional power supply. It is versatile, owing to its ease of resonant frequency tunability through multiple degrees of freedom in the design parameters of the mass-decorated membrane. The vibration sensing technology in this paper has the advantages of both contact-based and non-contact-based methods to satisfy the requirement for the future vibration sensing technology.

## Results

**Design and characterization of the vibration processing device**

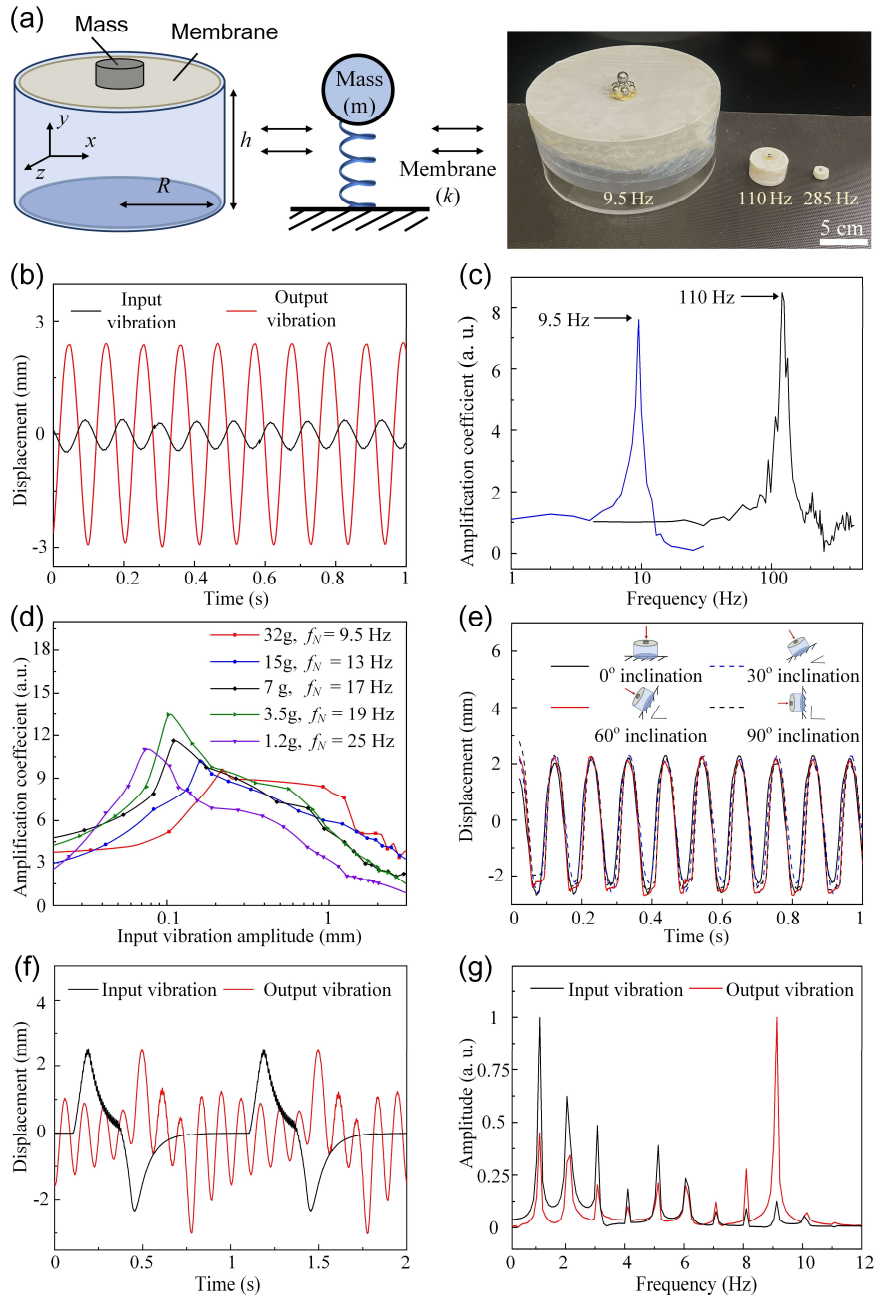

**Fig. 2. Characterization of the device.** (a) Device schematics and a photo of fabricated devices with different resonance frequencies (9.5 Hz, 110 Hz and 285 Hz). (b) Input and output vibration waveforms when the input frequency is 9.5 Hz measured from the device with 9.5 Hz eigenfrequency. (c) Amplification coefficients as a function of input frequency for two devices with 9.5 Hz and 110 Hz eigenfrequency respectively. (d) Amplification coefficient as a function of input vibration amplitude for devices with different loaded mass *m*. (e) Output waveforms when the device has different inclination angles with the ground. (f) Output vibration waveform when the input waveform is non-sinusoidal. (g) Frequency spectrums of the waveforms in (f).

As shown in Fig. 2(a), the vibration amplification device is composed of a mass-loaded membrane mounted on the top surface of a rigid cylindrical shell. The bottom of the shell is attached to the vibrating surface. The system can be characterized in a simplified form as the combination of a mass $m$ and a spring with stiffness $k$ determined by the membrane size, thickness, material, and strain. The vibration of the rigid cylinder follows the motion of the vibrating subject, which is the input of the device. The output is the vibration of the mass-loaded membrane. The mass-loaded membrane can be excited to resonance when the input vibration has a frequency close to its eigenfrequency. At this frequency, the vibration of the membrane will be largely enhanced compared to the vibration of the bottom surface. The eigenfrequency of the device can be adjusted to different values by changing one or multiple design parameters including the size, thickness, and material of the membrane, as well as weight and location of the loaded mass. In the Supplementary Fig. S2, we provide a systematic study of the tunable range of the eigenfrequency as a function of the above parameters.

In device characterization, the bottom of the device is attached to a rigid plate. The vibration is triggered by exciting a shaker (Bruel & Kjaer Type 4809) connected to the rigid plate. The vibration of both the bottom plate and the top membrane are measured with a laser vibrometer (Polytec PSV-200 Scanning Laser Vibrometer System) for a range of frequencies. Fig. 2(b) shows an example set of input and output vibration signals measured from a sample device with eigenfrequency of 9.5 Hz. When the input vibration has a frequency of 9.5 Hz and vibration amplitude of 0.76 mm, the output vibration amplitude reaches 5.3 cm, showing approximately 7-fold amplification. We define the peak-peak displacement of the membrane normalized by the peak-peak displacement of the bottom plate as the amplification coefficient. Fig. 2(c) shows the experimentally measured amplification coefficient for two different devices having the eigenfrequency of 9.5 Hz and 110 Hz, respectively. For each device, there is a peak in the amplification coefficient vs. frequency curve near its eigenfrequency, with approximately 7-fold and 9-fold vibration amplitude amplifications, respectively. When the input frequency is smaller than the eigenfrequency, the amplification coefficient is approximately one, meaning that there is neither attenuation nor amplification. The amplification coefficient drops quickly when the vibration frequency gets larger than the eigenfrequency. In Fig. 2(d), we demonstrate the amplification coefficient under different loaded mass and input vibration amplitude. When the loaded mass varies, the eigenfrequency as well as the corresponding peak amplification coefficient achieved at that frequency changes. As shown in Fig. 2(d), as the loaded mass changes from 1.2 g to 32 g, the eigenfrequency varies from 25 Hz to 9.5 Hz. The maximum amplification coefficient of more than 12 is observed at a mass load of 3.5 g. For a fixed mass load, when the input vibration amplitude increases from zero, the maximum amplification near the eigenfrequency first increases and then decreases, showing the nonlinearity of the device. Besides that, the device is tested when attached to a vibrating surface with different inclination angles to the ground, as illustrated in Fig. 2(e). With the same input excitation signal, the measured displacement of the top membrane layer is similar for different inclination angles, showing that the device is robust under inclination.

**Analogue vibration signal filtering effect of the proposed device**

The device can also be considered as an analog vibration signal filter. When the input vibration waveform is more complex than a single-frequency sinusoidal wave, such as the waveforms of human breaths, heartbeats, and other more realistic vibrations, part of the frequencies will be amplified, and part will be diminished. Fig. 2(f) shows a triangle-like input non-sinusoidal vibration waveform with a fundamental frequency of 1Hz and a peak-to-peak amplitude of 5 mm, provided by the shaker and measured by the laser vibrometer. Its frequency spectrum (black curve in Fig. 2(f)) contains multiple peaks centered at 1 Hz, 2 Hz, 3 Hz, etc. with decreasing amplitude. Upon filtering by the device with an eigenfrequency of 9.5 Hz, the peak near the eigenfrequency of the device is largely amplified. We observed the change of the frequency peak around the device's eigenfrequency for several different input waveforms with different fundamental frequencies either smaller or larger than 9.5 Hz (Fig. S3). This newly adapted frequency peak can be exploited as a label for the device. When there are multiple vibration sources in the sensing environment, for example in remote heartbeat/breath sensing when there are multiple persons in the same room, it was difficult to separate the signals from different persons in existing Wi-Fi remote sensing technologies[46]. We can attach devices with different eigenfrequencies to different vibration sources and use the eigenfrequency labels to differentiate the signals from each device. Similarly, in an industrial setting, the eigenfrequency of the passive device could be used to distinguish one machine from another. In addition, the device opens the possibility of analog vibration filtering and processing that have the potential to reduce digital processing burden and power consumption, improve signal to noise ratio, and enhance system security[47,48].

**Improved sensitivity and labeling effect in wireless Wi-Fi vibration sensing**

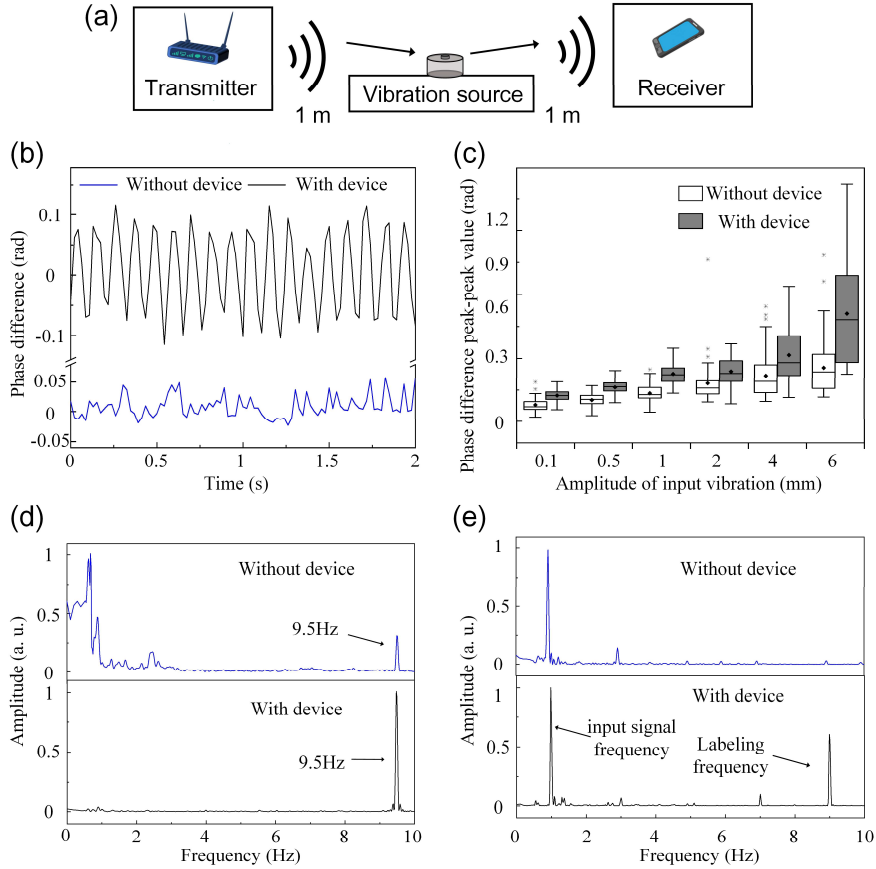

**Fig. 3. The application of fabricated device in Non-contact Wi-Fi vibration detection.** (a) A schematic illustration of the device application in Wi-Fi vibration detection, (b) Experimentally measured CSI phase difference between the two receiver antennas listening on the same Wi-Fi channel. The input vibration amplitude is set as 1 mm. (c) Statistic box plot (displays the minimum, first quartile, median, third quartile, and the maximum as a box, mean as a dot in box, and outlier as dots outside the box) of the experimental CSI phase difference peak-peak values for 52 Wi-Fi channels. The frequency spectrum of the vibration signal measured by Wi-Fi for (d) amplification (the input vibration is 9.5 Hz sinusoidal wave with amplitude of 1mm) and (e) labeling (the input signal is non-sinusoidal with amplitude of 5 mm (the same as the one in Fig. 2(f)).

Wi-Fi is one of the most common wireless communication technologies to transfer data between two devices. In the recent past, it is being considered as a potential tool for passive sensing of the surrounding environment due to the ubiquitous presence of Wi-Fi devices. Particularly, it has been explored for its potential in contactless vibration sensing[26]. Wi-Fi based vibration sensing utilizes a Wi-Fi router emitting a signal, which is reflected by a vibration source and captured at the receiver, for example, a cellphone or a Wi-Fi router. The receiver then calculates the Channel State Information (CSI), which captures the frequency response of the wireless channel between the Wi-Fi transmitter and the receiver. An accurate estimate of the CSI has been shown to be highly sensitive to changes in the environment[49]. Therefore, CSI is used to calculate the vibration frequency of the surface of interest that is placed between the Wi-Fi transmitter and receiver[27]. However, current Wi-Fi vibration sensing methods are not widely used in practice due to their limited sensitivity. For example, its lack of sensitivity to small vibration amplitudes and its inability to differentiate between different objects limits its practical usability. In our proposed system, our passive vibration processing device aims to overcome the sensitivity limitation of Wi-Fi based sensing by amplifying the vibration signal and in turn its impact on CSI measurement at the Wi-Fi receiver. The ability of our proposed device to assign a unique eigenfrequency to identify an object in an environment addresses the second limitation; thus, we leverage the benefits of low-cost, non-contact Wi-Fi based sensing, as well as overcome its limitations using our novel design of the device.

Consider the device with a 9.5 Hz eigenfrequency attached to the surface of a vibrating sample. It amplifies the movement (vibration) amplitude of the top membrane compared to the input vibration of the bottom layer, resulting in a more apparent change in the reflected Wi-Fi signal and its estimated CSI, which is then used by the Wi-Fi receiver to accurately determine the vibrating frequency. The methodology for Wi-Fi sensing used in this paper is detailed in the methods section. The frequency of Wi-Fi operation used in this study is 5 GHz with a channel bandwidth of 20 MHz, containing 52 Wi-Fi channels for measurement (more details of the hardware and settings in the Methods section). In Fig. 3(b), we plot the CSI phase difference between two antennas at the Wi-Fi receiver, with and without the amplification device for a single Wi-Fi channel. The phase difference of CSI is a function of the displacement of the vibrating surface. The Fourier frequency spectrum of the CSI phase difference indicates the vibration frequencies and their strength. The variations of CSI phase difference peak-to-peak value, measured over six observations of 9.5 Hz vibrations with vibrating amplitude ranging from 0.1 mm to 6 mm is plotted in Fig. 3(c). In the presence of our proposed device, the peak-to-peak amplitude of CSI phase difference is improved even at extremely small vibration amplitude of 0.1 mm. An instance of the frequency spectrum obtained from the CSI phase difference with an input vibration amplitude of 1 mm is shown in Fig. 3(d). In the absence of a device attached to the vibration source, the amplitude of the 9.5 Hz peak in the vibration spectrum is lower than that of frequencies corresponding to environmental noise; this low amplitude is below the measurement limit of our Wi-Fi system, resulting in poor performance in detecting vibrations. Upon attachment of the device, the 9.5 Hz peak amplitude is amplified, and the Wi-Fi system accurately measures the real vibration frequency. The accuracy for different vibration amplitude (measured by the correct rate of Wi-Fi sensing dominant frequency results among different time periods) is also improved with the device, as shown in Fig. S4. With the device, vibration amplitudes as low as 1 mm are detected with 100% accuracy, due to its amplification. In contrast, without the device, even vibration amplitudes of 4 mm go unnoticed by the Wi-Fi receiver. Furthermore, we demonstrate the labeling function of the device in Wi-Fi vibration sensing. The signal shown in Fig. 2(f) is used as the input and the measured vibration frequency spectrum is shown in Fig. 3(e). Apart from the input frequency 1 Hz, a 9 Hz labeling frequency peak is also observed, which can be used to label and distinguish the vibration surface.

**Passive sensitivity enhancement of camera-based non-contact vibration measurement**

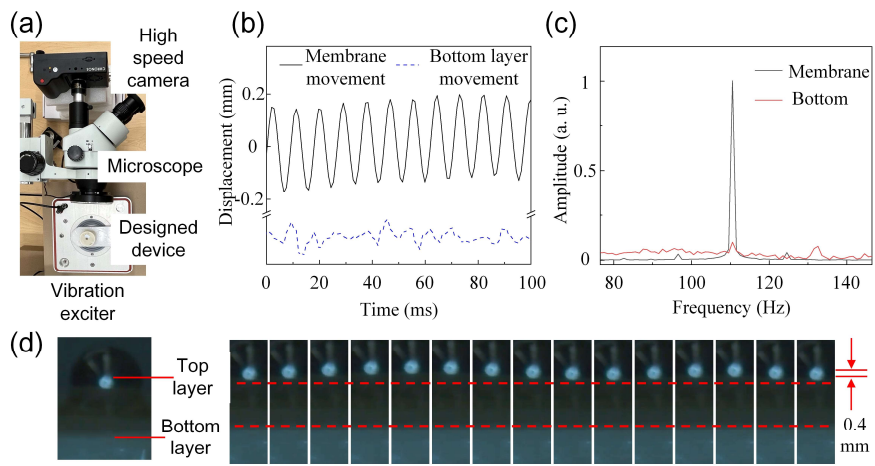

**Fig. 4. Vibration amplification device for camera-based vision method.** (a) Experiment setup. (b) The extracted vibration waveforms from video S1 (the eigenfrequency of the device is 110 Hz). (c) Frequency spectrums of the vibration waveforms in Fig. 4(b). (d) Screenshots of the vibration video at different times (interval = 1 ms).

As digital cameras become widely accessible, camera-based detection and monitoring techniques are quickly developed. As a second example of application, we show the device's performance in improving the sensitivity of vibration sensing using a camera (Chronos 2.1-HD). The experiment setup is shown in Fig. 4(a). The input vibration frequency and the device eigenfrequency are both set to 110 Hz. The device is mounted on a rigid plate on top of a shaker. To obtain an image with better resolution, a microscope is used in combination with the high-speed camera to record the vibration video of the device. An example video is shown in Video S1. The displacement of the sample can be extracted by image processing methods. Here we first make an object segmentation and then analyze the movement of the top membrane layer and bottom layer[50,51] (details are described in Methods). The extracted displacement waveform is plotted in Fig. 4(b), where the membrane layer has a more enhanced movement than the bottom layer. The input and output vibration waveforms are plotted in Fig. 4(b). The peak-to-peak amplitudes of input and output vibrations are around 0.1 mm and 0.4 mm, respectively, showing a 4-fold amplification. The frequency spectrums of the input and output vibrations are shown in Fig. 4(c). Compared with the bottom layer movement, the top layer has a clear peak at 110 Hz, demonstrating the enhancement of the vibration. The snapshots of the video are shown in Fig. 4(d), demonstrating the amplified vibration of the mass on the membrane. Furthermore, we also fabricated another device with a similar structure but having a higher eigenfrequency of 285 Hz. The vibration waveforms are extracted from the recorded video (see Supplementary Video S2 and S3) by the pixel value changes on the vibration sample boundary and vibration amplitude enhancement of 8.4 folds after applying the device is also observed (details are presented in Supplementary Materials Fig. S6).

**Multifunctional editing of the device**

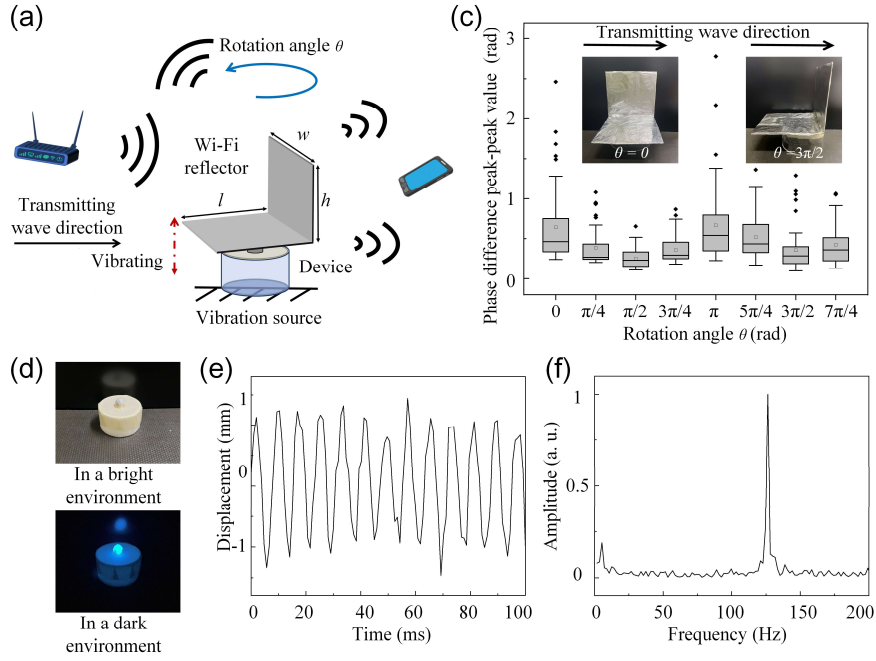

**Fig. 5. Multifunctional editing of the device: directional Wi-Fi reflector and fluorescent central load.** (a) Demonstration of the designed Wi-Fi reflector mounted on the top of the device in Wi-Fi sensing and the control of wave by different rotation angle. (b) Experiment setup demonstration of the of the eight experiments with different top layer rotation angle. (c) Box plot of the measured CSI phase difference peak-peak value of 52 Wi-Fi channels with eight different reflector rotation angles. (d) the pictures of fluorescent layer coated device in bright environment and dark environment (the corresponding video is Supplementary S4 and S5, respectively). (e) The extracted vibration waveform from Supplementary Video S5 of fluorescent-coated device in dark environment. (f) Frequency spectrum of Fig. 5(e).

Apart from the amplification and filtering functions, the device also provides a platform for surface editing that can add more functions to the sensing system. In this paper, we demonstrate two examples of surface editing.

For the Wi-Fi-based vibration sensing, a directional reflector (shown in Fig. 5(a)) is attached to the top membrane of the device to replace the loaded mass in the original design. The new device has an eigenfrequency of 5 Hz. The reflector is composed of a vertical plate connected to a horizontal plate, which can modify the directivity and the far-field radiation pattern of the reflected Wi-Fi signal[52]. This influence about wave intensity on the direction can be used in Wi-Fi vibration sensing. In the experiment using the reflector, the input vibration frequency is at the eigenfrequency of 5 Hz. The position of Wi-Fi transmitter and receiver are fixed and we rotate the reflector to different orientations with the rotation angle $\theta$. At each orientation, the measured peak-to-peak phase difference between two antennas across all 52 Wi-Fi channels is shown in Fig. 5(c). The results have clear angle-dependence. The first and fifth orientations have larger phase differences due to the reflector directivity. Our Wi-Fi signal reflector example provides the possibility of surface modifications to improve wireless reception quality, reduce energy consumption, and achieve better security and privacy.

For the camera vibration sensing, a fluorescent layer is coated on the surface of the loaded mass in the device with eigenfrequency of 125 Hz. The input vibration has a frequency of 125 Hz. The pictures of the coated device in bright and dark environments are shown in Fig. 5(d). Because of the fluorescent coating, the vibration of the device can still be captured by a camera in dark environments, as demonstrated in Fig. 5(e) and (f), as well as the recorded video (see Video S4 and Video S5).

## Discussion

This study presents a passive mechanical vibration processing device mounted on the vibration surface for real-time, low-cost, and wireless vibration sensing. The device performs analog amplification and labeling on vibration signals, which are used in Wi-Fi and camera-based vibration detection. Most of the existing works on non-contact vibration measurement have focused on software or signal processing approaches on the data acquired[23,46]. Our design of a hybrid sensing system, which leverages a passive physical device attached to the vibrating surface, improves the accuracy of the sensed data physically thus reduces the overhead on digital signal processing needs. The proposed device is composed of a mass-loaded membrane, whose vibration eigenfrequency demonstrated in this work are 9.5 Hz, 110 Hz and 285 Hz. It is relatively straight forward to expand the range of eigenfrequencies to be lower than 1 Hz or to as high as few MHz by modifying the device dimensions and the fabrication methods, in order to meet the working demands of various applications.

We demonstrated up to 13.35 folds of vibration amplitude amplification. The amplification coefficient can be further improved by using optimizing the combination of membrane material properties, membrane size, and the loaded mass, as well as by choosing a membrane material with lower vibration damping. The direct application of our device is to improve the sensitivity of non-contact vibration sensing systems, such as camera-based or Wi-Fi/RFID based vibration sensing systems. In this work, we improved the sensitivity of Wi-Fi and camera vibration sensing by around one order of magnitude. It is well known that high frequency vibrations usually have smaller vibration amplitudes[53], limiting the highest detectable frequency by vibration sensing systems. With the help of our vibration amplitude amplification device, the maximum detectable vibration frequency range can be expanded.

Our device serves as a physical layer filter for the vibration signals. In this paper, we demonstrated that a frequency peak near the first eigenfrequency of the device will always appear in the spectrum of the output vibrations. This frequency peak can be used as a label to address another major challenge in wireless vibration measurement: source identification. It is difficult to distinguish two vibrating sources using state-of-the-art wireless sensing techniques. Our proposed device addresses this problem by assigning a unique eigenfrequency to the device that is attached to each source in an environment. For example, in structure health monitoring applications, multiple devices can be attached to different positions of the structure to sense the location-dependent vibration mode changes caused by structure damage. Beyond the labeling function, more filtering functions can be developed in the future by manipulating the frequency response of the device. For example, the higher order eigenmodes that have not been considered in this paper have richer properties to be explored for analog vibration signal processing.

Multifunctional editing of the device's top layer will add more functions to wireless vibration sensing systems. In a Wi-Fi based system, we integrated a directional Wi-Fi reflector to our device to provide direction-dependent sensitivity. This additional reflector can be used to improve wireless reception quality and achieve better security. In a camera-based system, we added fluorescent material to achieve vibration detection in a dark environment. The flexibility of our device in terms of its shape, size, and materials render it as a suitable platform for surface editing for multi-functional vibration sensing.

In conclusion, our proposed device will make wireless vibration sensing more accessible and benefit a wide range of applications in personalized healthcare, industrial structure inspection, natural disaster monitoring, and material characterization, and promote the integration of sensing capabilities into the rapid development of wireless communication and cloud computing infrastructures.

## Materials and Methods

### Device Fabrication and laser vibrometer measurement setup

The membrane of the device is made of rubber thin film (McMaster-Carr Supply Company), and the mass loads at the center of the membrane are steel balls. The cylindrical hard shell of device with 9.5 Hz eigenfrequency is composed of acrylic plastic components and the shells of other smaller devices are fabricated by 3D printing. The detailed parameters of the designed devices and their corresponding eigenfrequencies can be found in Supplementary Table S1. The directional reflector is made of hard cardboard covered with tinfoil. For the fluorescent coating layer, fluorescence particles (SEISSO company) are added to the silicone rubber solution (PIXISS company) and coated on the surface of the steel ball that is used as the loaded mass of the device.

The experimental setup for characterizing the devices' vibration properties can be viewed in Supplementary Fig. S1. The input vibration waveform is generated from a function generator (Tektronix AFG3022C) and then input to a power amplifier. The signal from the power amplifier controls a shaker (Bruel & Kjaer Type 4809) to provide the input vibration to the hard plate, where the device is attached. A laser vibrometer system (Polytec OFV 056 with scanning head PSV-200) is used to measure the transverse vibration motion of the mass-loaded membrane when device is attached, and the supporting plate motions without the device.

**Wi-Fi measurement setup**

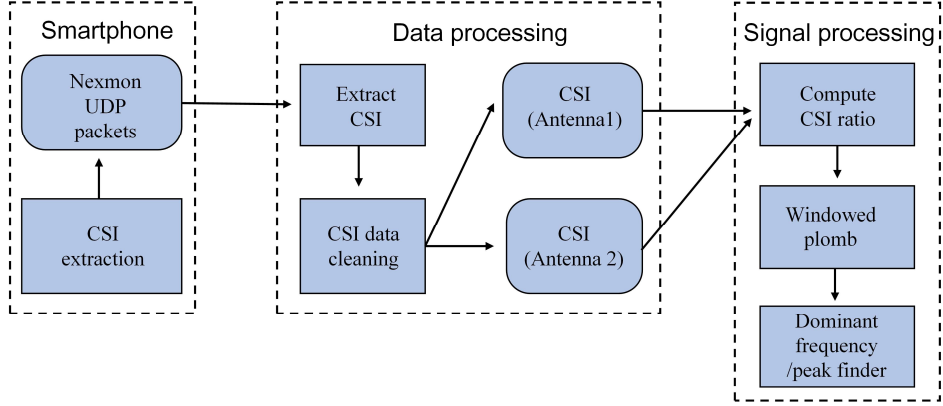

Fig. 6. An overview of the Wi-Fi Sensing System

The CSI in Wi-Fi sensing is mathematically represented as the superposition of a signal from all the N paths as shown in Eq. 1

$$H(f;t) = \sum_{n}^{N} a_n(t) e^{-j2\pi f \tau_n(t)} \tag{1}$$

where $a_n(t)$ represents the amplitude attenuation factor, $\tau_n(t)$ is the phase of the electromagnetic signal due to the propagation delay (imaginary component of a particular path), and $f$ is the carrier frequency. The CSI amplitude ($|H|$) and phase ($\angle H$) are impacted by the displacements and movements of the transmitter, receiver, and surrounding objects and humans. Therefore, CSI effectively represents the medium of signal propagation and any changes to CSI in a stable environment resulting from movement of an individual or the device in this experiment can be detected. The device movement results in periodic changes to the phase and should therefore be the dominant frequency in the frequency response of the phase difference signal measured across time.

In this study, the commercial Asus RT-AC86U Wi-Fi router is used as the transmitter and captured by a rooted smartphone (Nexus 6P) with bcm4358 Wi-Fi chip. The Wi-Fi beacon interval is set to 20 ms and distances between router and measured sample, sample and smartphone are both set as 1m. The measured receiving signal strength is −25 to −30 dB and the time duration of one vibration measurement is set as 30 s. As Wi-Fi 4 or 802.11n utilizes OFDM (Orthogonal Frequency Division Multiplexing) for transmission, 52 sub-carrier (or channels) CSI across 20 MHz can be computed. The Nexmon CSI tool is used to extract the CSI at the receiver for each antenna. We perform basic preprocessing such as angle unwrap and time-based correlation between the packets received on each antenna at the receiver[49].

CSI phase difference between two antennas at the receiver is then calculated and used to determine the amplitude and frequency of vibration of the surface of interest. As the signal is measured irregularly, we apply plomb, a MATLAB based implementation of the Lomb–Scargle Periodogram[54,55], to obtain the Power Spectral Density (PSD), which is further analyzed to obtain the dominant frequency in a given frequency range. The whole procedure of the dominant frequency calculation can be viewed in Supplementary Algorithm Table S2. The computations are performed using a sliding window protocol to limit the time frame, over which the plomb is applied to improve its accuracy. The time window is set as 15 s and the window slides forward by 20 ms. The accuracy results shown in Fig. S4 is averaged over a large number of time windows getting the right vibration frequency and the total time window number is used to calculate accuracy.

**Video recording and moving trajectory extraction**

The video recording system for camera-based vibration sensing is shown in Fig. 4(a). To achieve better video resolution, a microscope with 5 times amplifications is used when recording Supplementary Video S1 while 20 times amplifications for video S2 and video S3. Video S4 and S5 are recorded without the microscope. The camera (Chronos 2.1-HD) for recording the videos has frame per second (fps) of 1069.61. The camera lens is set to be parallel to the vibrating surface. Two video processing methods are used to extract the vibration waveform from the recorded videos. The first one is based on vibrating object segmentation and moving trajectory extraction. As shown in the supplementary Fig. S4(a-c), a snapshot is first converted to grayscale picture with value 0-255 to represent the color in every pixel. The moving sample is then segmented by a threshold and the centroid of it can be calculated. Movement of this centroid with time (shown in Fig. S4(d)) can be used to analyze the vibration frequency. This method is used for the 110 Hz device. Furthermore, the grayscale value changes on the vibration sample boundary are used to acquire vibration frequency for the 285 Hz device, which can deal with the smaller vibration movement compared to the first method [56]. The edge of the moving sample (where the grayscale value has the largest variation) is first obtained for both top layer central mass and bottom hard shell. Points on these edges are chosen and the color value changes with time at these points are plotted in Fig. S5(c) and (d).

# References

1. Newland, D. E. & Ungar, E. E. Mechanical Vibration Analysis and Computation. *The Journal of the Acoustical Society of America* **88**, 2 (1990).

2. Mühlhans, J. H. Low frequency and infrasound: A critical review of the myths, misbeliefs and their relevance to music perception research. *Musicae Scientiae* **21**, 267–286 (2017).

3. Liu, J. *et al.* Tracking Vital Signs During Sleep Leveraging Off-the-shelf WiFi. in *Proceedings of the 16th ACM International Symposium on Mobile Ad Hoc Networking and Computing* 267–276 (ACM, 2015). doi:10.1145/2746285.2746303.

4. Doebling, S. W., Farrar, C. R. & Prime, M. B. A Summary Review of Vibration-Based Damage Identification Methods. *The Shock and Vibration Digest* **30**, 91–105 (1998).

5. Henriquez, P., Alonso, J. B., Ferrer, M. A. & Travieso, C. M. Review of Automatic Fault Diagnosis Systems Using Audio and Vibration Signals. *IEEE Trans. Syst. Man Cybern, Syst.* **44**, 642–652 (2014).

6. Shifat, T. A. & Hur, J. W. An Effective Stator Fault Diagnosis Framework of BLDC Motor Based on Vibration and Current Signals. *IEEE Access* **8**, 106968–106981 (2020).

7. Jiang, S., Xie, H., Shan, J. & Mak, K. F. Exchange magnetostriction in two-dimensional antiferromagnets. *Nat. Mater.* **19**, 1295–1299 (2020).

8. Gavartin, E. A hybrid on-chip optomechanical transducer for ultrasensitive force measurements. *Nat. Nanotechnol* **7**, 6 (2012).

9. Barnard, A. W., Zhang, M., Wiederhecker, G. S., Lipson, M. & McEuen, P. L. Real-time vibrations of a carbon nanotube. *Nature* **566**, 89–93 (2019).

10. Gil-Santos, E. *et al.* Optomechanical detection of vibration modes of a single bacterium. *Nat. Nanotechnol.* **15**, 469–474 (2020).

11. Jiang, T., Li, C., He, Q. & Peng, Z.-K. Randomized resonant metamaterials for single-sensor identification of elastic vibrations. *Nat Commun* **11**, 2353 (2020).

12. Zeng, Z. *et al.* Sustainable-Macromolecule-Assisted Preparation of Cross-linked, Ultralight, Flexible Graphene Aerogel Sensors toward Low-Frequency Strain/Pressure to High-Frequency Vibration Sensing. *Small* **18**, 2202047 (2022).

13. Marques dos Santos, F. L., Peeters, B., Lau, J., Desmet, W. & Goes, L. C. S. The use of strain gauges in vibration-based damage detection. *J. Phys.: Conf. Ser.* **628**, 012119 (2015).

14. Watakabe, M., Itoh, Y., Mita, K. & Akataki, K. Technical aspects of mechnomyography recording with piezoelectric contact sensor. *Med. Biol. Eng. Comput.* **36**, 557–561 (1998).

15. Yaghootkar, B., Azimi, S. & Bahreyni, B. A High-Performance Piezoelectric Vibration Sensor. *IEEE Sensors J.* **17**, 4005–4012 (2017).

16. Delgado, A. V., González-Caballero, F., Hunter, R. J., Koopal, L. K. & Lyklema, J. Measurement and interpretation of electrokinetic phenomena. *Journal of Colloid and Interface Science* **309**, 194–224 (2007).

17. Lee, J. J. & Shinozuka, M. Real-Time Displacement Measurement of a Flexible Bridge Using Digital Image Processing Techniques. *Exp Mech* **46**, 105–114 (2006).

18. Devillez, A. & Dudzinski, D. Tool vibration detection with eddy current sensors in machining process and computation of stability lobes using fuzzy classifiers. *Mechanical Systems and Signal Processing* **21**, 441–456 (2007).

19. Lawson, C. & Ivey, P. Tubomachinery blade vibration amplitude measurement through tip timing with capacitance tip clearance probes. *Sensors and Actuators A: Physical* **118**, 14–24 (2005).

20. Nassif, H. H., Gindy, M. & Davis, J. Comparison of laser Doppler vibrometer with contact sensors for monitoring bridge deflection and vibration. *NDT & E International* **38**, 213–218 (2005).

21. Son, K.-S., Jeon, H.-S., Chae, G.-S., Park, J.-S. & Kim, S.-O. A fast high-resolution vibration measurement method based on vision technology for structures. *Nuclear Engineering and Technology* **53**, 294–303 (2021).

22. Chen, J. G. *et al.* Modal identification of simple structures with high-speed video using motion magnification. *Journal of Sound and Vibration* **345**, 58–71 (2015).

23. Zona, A. Vision-Based Vibration Monitoring of Structures and Infrastructures: An Overview of Recent Applications. *Infrastructures* **6**, 4 (2020).

24. Ambrosanio, M., Franceschini, S., Grassini, G. & Baselice, F. A Multi-Channel Ultrasound System for Non-Contact Heart Rate Monitoring. *IEEE Sensors J.* **20**, 2064–2074 (2020).

25. Yu Gu, Xiang Zhang, Zhi Liu, Fuji Ren. WiFi-based Real-time Breathing and Heart Rate Monitoring during Sleep. *2019 IEEE Global Communications Conference (GLOBECOM)* (2019).

26. Wu, C., Wang, B., Au, O. C. & Liu, K. J. R. Wi-Fi Can Do More: Toward Ubiquitous Wireless Sensing. *IEEE Comm. Stand. Mag.* **6**, 42–49 (2022).

27. Zeng, Y. *et al.* FarSense: Pushing the Range Limit of WiFi-based Respiration Sensing with CSI Ratio of Two Antennas. *Proc. ACM Interact. Mob. Wearable Ubiquitous Technol.* **3**, 1–26 (2019).

28. Li, L., Xiaoguang, H., Ke, C. & Ketai, H. The applications of WiFi-based Wireless Sensor Network in Internet of Things and Smart Grid. in *2011 6th IEEE Conference on Industrial Electronics and Applications* 789–793 (IEEE, 2011). doi:10.1109/ICIEA.2011.5975693.

29. Yang, Z., Mei, J., Yang, M., Chan, N. H. & Sheng, P. Membrane-Type Acoustic Metamaterial with Negative Dynamic Mass. *Phys. Rev. Lett.* **101**, 204301 (2008).

30. Sun, L. *et al.* Membrane-type resonator as an effective miniaturized tuned vibration mass damper. *AIP Advances* **6**, 085212 (2016).

31. Lu, Z., Yu, X., Lau, S.-K., Khoo, B. C. & Cui, F. Membrane-type acoustic metamaterial with eccentric masses for broadband sound isolation. *Applied Acoustics* **157**, 107003 (2020).

32. Dong, L., Grissom, M., T. Fisher, F., & 1 Department of Mechanical Engineering, Stevens Institute of Technology, Hoboken, NJ 07030, USA; Resonant frequency of mass-loaded membranes for vibration energy harvesting applications. *AIMS Energy* **3**, 344–359 (2015).

33. Huang, J. *et al.* Vibration monitoring based on flexible multi-walled carbon nanotube/polydimethylsiloxane film sensor and the application on motion signal acquisition. *Nanotechnology* **31**, 335504 (2020).

34. Wang, Y., Wang, W., Zhou, M., Ren, A. & Tian, Z. Remote Monitoring of Human Vital Signs Based on 77-GHz mm-Wave FMCW Radar. *Sensors* **20**, 2999 (2020).

35. Z. Kotta, H., Rantelobo, K., Tena, S. & Klau, G. Wireless Sensor Network for Landslide Monitoring in Nusa Tenggara Timur. *TELKOMNIKA* **9**, 9 (2011).

36. Farrar, C. R., Darling, T. W., Migliori, A. & Baker, W. E. MICROWAVE INTERFEROMETERS FOR NON-CONTACT VIBRATION MEASUREMENTS ON LARGE STRUCTURES. *Mechanical Systems and Signal Processing* **13**, 241–253 (1999).

37. Whitlow, R. D. *et al.* Remote Bridge Monitoring Using Infrasound. *J. Bridge Eng.* **24**, 04019023 (2019).

38. Moschas, F. & Stiros, S. Measurement of the dynamic displacements and of the modal frequencies of a short-span pedestrian bridge using GPS and an accelerometer. *Engineering Structures* **33**, 10–17 (2011).

39. Xia, H., De Roeck, G., Zhang, N. & Maeck, J. Experimental analysis of a high-speed railway bridge under Thalys trains. *Journal of Sound and Vibration* **268**, 103–113 (2003).

40. Gentile, C. & Bernardini, G. An interferometric radar for non-contact measurement of deflections on civil engineering structures: laboratory and full-scale tests. *Structure and Infrastructure Engineering* **6**, 521–534 (2010).

41. Chen, Z., Liu, J., Zhan, C., He, J. & Wang, W. Reconstructed Order Analysis-Based Vibration Monitoring under Variable Rotation Speed by Using Multiple Blade Tip-Timing Sensors. *Sensors* **18**, 3235 (2018).

42. Zhao, L., Huang, X., Zhao, Y. & Si, W. Design of a wireless vibration metre for conductor vibration monitoring. *Struct Control Health Monit* **25**, e2143 (2018).

43. Zalalutdinov, M. K. *et al.* Acoustic cavities in 2D heterostructures. *Nat Commun* **12**, 3267 (2021).

44. Langfeldt, F., Riecken, J., Gleine, W. & von Estorff, O. A membrane-type acoustic metamaterial with adjustable acoustic properties. *Journal of Sound and Vibration* **373**, 1–18 (2016).

45. Yildirim, T., Ghayesh, M. H., Li, W. & Alici, G. A review on performance enhancement techniques for ambient vibration energy harvesters. *Renewable and Sustainable Energy Reviews* **71**, 435–449 (2017).

46. Chen, C., Han, Y., Chen, Y. & Liu, K. J. R. Multi-person breathing rate estimation using time-reversal on WiFi platforms. in *2016 IEEE Global Conference on Signal and Information Processing (GlobalSIP)* 1059–1063 (IEEE, 2016). doi:10.1109/GlobalSIP.2016.7906004.

47. Caloz, C., Gupta, S., Zhang, Q. & Nikfal, B. Analog Signal Processing: A Possible Alternative or Complement to Dominantly Digital Radio Schemes. *IEEE Microwave* **14**, 87–103 (2013).

48. Pors, A., Nielsen, M. G. & Bozhevolnyi, S. I. Analog Computing Using Reflective Plasmonic Metasurfaces. *Nano Lett.* **15**, 791–797 (2015).

49. Gringoli, F., Schulz, M., Link, J. & Hollick, M. Free Your CSI: A Channel State Information Extraction Platform For Modern Wi-Fi Chipsets. in *Proceedings of the 13th International Workshop* *on Wireless Network Testbeds, Experimental Evaluation & Characterization - WiNTECH '19* 21–28 (ACM Press, 2019). doi:10.1145/3349623.3355477.

50. Chen, L., Shen, J., Wang, W. & Ni, B. Video Object Segmentation Via Dense Trajectories. *IEEE Trans. Multimedia* **17**, 2225–2234 (2015).

51. Fragkiadaki, K., Geng Zhang, & Jianbo Shi. Video segmentation by tracing discontinuities in a trajectory embedding. in *2012 IEEE Conference on Computer Vision and Pattern Recognition* 1846–1853 (IEEE, 2012). doi:10.1109/CVPR.2012.6247883.

52. Chan, J., Zheng, C. & Zhou, X. 3D Printing Your Wireless Coverage. in *Proceedings of the 2nd International Workshop on Hot Topics in Wireless* 1–5 (ACM, 2015). doi:10.1145/2799650.2799653.

53. Shao, L. *et al.* Femtometer-amplitude imaging of coherent super high frequency vibrations in micromechanical resonators. *Nat Commun* **13**, 694 (2022).

54. Lomb, N. R. Least-squares frequency analysis of unequally spaced data. *Astrophys Space Sci* **39**, 447–462 (1976).

55. Scargle, J. D. Studies in astronomical time series analysis. II - Statistical aspects of spectral analysis of unevenly spaced data. *ApJ* **263**, 835 (1982).

56. Son, K.-S., Jeon, H.-S., Park, J.-H. & Park, J. W. Vibration displacement measurement technology for cylindrical structures using camera images. *Nuclear Engineering and Technology* **47**, 488–499 (2015).

## Acknowledgments

**Funding:** C.M. and D.Z. acknowledge the funding support from UW Madison startup package. B.K., A.P., and M.O.S. acknowledge the funding support the following NSF grants: CCSS-2034415, CNS - 2107060, CNS-2142978, CNS-2213688, CNS-2112562, and Wisconsin Alumni Research Foundation awards.

**Author contributions:** D.Z., C.M., and B.K. conceived the research idea. D.Z. performed the mechanical simulation and fabricated the devices. D.Z., A.P., and M.O.S.

performed the experiments, collected the data, and did the analysis. D.Z., C.M. A.P. and B.K. contributed to the paper writing. C.M. supervised the study.

**Competing interests:** Authors declare that they have no competing interests.

**Data and materials availability:** All data needed to evaluate the conclusions in the paper are present in the paper and/or the Supplementary Materials. Additional data related to this paper may be requested from the authors.